\documentclass[10pt]{article}
\usepackage{amssymb,amsmath}
\usepackage[dvips,unicode]{hyperref}
\usepackage[dvips]{graphicx}

\begin{document}

\title {Exotic decay modes of the nucleon resonance $N^*(1535)$
and the $\theta^+$ problem}
\author {L. S. Molchatsky \thanks{Department of Theoretical Physics,
Samara State Pedagogical University, M.~Gor'ky Street 65/67,
443099 Samara, Russian Federation;}
\thanks{ e-mail:\,levmol@yandex.ru}}

\date{}
\maketitle

\begin{abstract} \noindent
 Baryon resonance state with strangeness $S = +1$ observed in systems
 $p\,K^0$ and   $n\,K^+$ is interpreted as a result of the decay $N^{*+} \to p \,\eta$
  of the well-known   resonance $N^*(1535)$ (with spin-parity $j^P = \frac12^-$
  and mass $m \approx 1.54$ GeV )    and   the subsequent oscillation
  $\eta \rightleftarrows K^0$. It is argued that the latter is noticeably
  enhanced by the neighbourhood of the $\eta$ and $K^0$ levels and the gluon
  exchanges of the "penguin" form.
\end{abstract}


\section{Introduction}
\label{int}
 Nowadays, the status of the exotic baryon resonance $ \theta^+ $ with strangeness
 $ S = +1 $ and mass $ m = 1.54 \pm 0.01~ GeV $, observed  by several
 experimental groups~[1~-~4] in 2003, still remains uncertain: some
groups ~[5,~6] confirm the existence of this resonance while
others
 ~[7~-~9] find no signal from the ~$\theta^+$. Its basic properties (spin-parity and total width)
   is not established either in any direct experiment. However, the phase
analysis of the $K\,N$ ~scattering ~[10] imposes severe
restriction on width of the resonance:
 $ \Gamma < 1~ MeV $.~ This fact seems to be great of interest. Such a narrow width
  is at variance with its theoretical estimate in the framework of a
   Chiral Soliton Model [11] although the $ \theta^+ $ has been predicted
  [12] by  this very model.

    Taking  this situation into account we consider here possibility of interpreting
 the resonance in the systems $ pK^0 $  and  $ nK^+ $  outside the standard approach,
  involving the hypothesis on the exotic baryon resonance~$ \theta^+ $.

\section{Exotic decay modes}
\label{exot}
 It should be noted that only the upper limit for the value of the resonance width
 has been experimentally established and it turned out
 to be much too small compared to the theoretical predictions ~[10,~11].
  Therefore, it seems to be a plausible proposition that the $ p\,K^0 $  and  $ n\,K^+ $
  states arise as a consequence of some very rare processes. We assume that
 they are a result of the two-particle decay of the well-known nucleon resonance
 $N^*(1535)$ ~[13]  (with spin-parity  $ j^P = \frac{1}{2}^- $ and $ m = 1.54~ GeV $~ )~
 into proton  and $\eta$ meson and the subsequent oscillation $ \eta^0 \rightleftarrows K^0 $.
 Thus, the $p\,K^0$ state is to arise as the final one  in the transitions:
\begin{equation}
\label{1}
 N^{*+} \to p\,\eta \to p\,K^0 .
\end{equation}

This is the main process. As to the $n\,K^+$ state, it will be
efficiently produced by means of the final state interaction, i.e.
the strong interaction reaction $K^0 \,p \to K^+ \,n$, provided
that the process (1) occurs.\footnote {There is also another
channel for producing the $N \, K^+$ state: $N^{*+} \to n \,
\pi^+ \to n \, K^+$. Here the latter transition is due to the
$\pi^+ \rightleftarrows K^+$ oscillation. However, this process
turns out to be rather suppressed because of a large distance
between the $\pi^+$ and $K^+$ levels.}
 Taking this fact into consideration, we shall completely disregard
 the last channel and concentrate on the investigation of the process (1).
Certainly, considered decay modes should be solely
low-probability because  the $\eta \rightleftarrows K^0$
oscillation can go only via weak interaction.

According to model (1), the branching ratio of the decay into the
proton and $ K^0 $ meson is defined by the relation
\begin{equation}
\label{2}
 \Gamma(N^{*+} \to p\,K^0)/\Gamma_t
 = \left(\Gamma(N^{*+} \to p\,\eta)/\Gamma_t\right) \langle P(\eta \to K^0)\rangle.
\end{equation}
Here $ \langle P(\eta \to K^0)\rangle $ is the probability of
 the $ \eta \to K^0$ transition averaged over period of oscillation.
 Values of all quantities on the right-hand of equation (2),
 excepting $ \langle P(\eta \to K^0)\rangle $, are experimentally
 established. So, we should proceed to its determination.

\section{The $ \eta - K^0$ oscillation}
\label{osc}

 It is obvious that the $\eta - K^0$ transition is due
to the $W$ boson exchange complemented to the virtual gluon
processes. In this respect it is  similar to the non - leptonic
strange particle decays. Having taken this fact into account, we
investigated the $\eta - K^0$ transition problem in the framework
of the non-leptonic effective Lagrangian theory ~ [14,~ 15], i.
e. by using an approach  typical for a theory of the non-leptonic
decays. In accord with ~[14,~15], a complete non-leptonic
effective Lagrangian is given by the expression
\begin{equation}
\label{3} L = \sqrt 2 \, G \sin\theta \cos\theta \sum_{i=1}^6 c_i
\, O_i,
\end{equation}
where $G$ is the Fermi constant of weak interaction; $\theta$ is
the Cabbibo angle; $O_i$ are 4-fermion operators consisting of $u,
d$ and $s$ quark operators; $c_i$  are the coupling constants.
Here and further on we adhere to the method of calculation  and
notations given in ~[15].

Detailed analysis of the contributions to the $\eta - K^0$
amplitude gives evidence to the fact that there exists an effect
dominating explicitly among other factors. It arises from the
virtual exchanges which form process known as the process of the
"penguin" form.  The  problem considered deals with virtual
annihilation of the neutral meson into gluons and with the
following sequence of the s-channel transitions:
\begin{equation}
\label{4} d \, \tilde d  \to g \to  u \, \tilde u \to d\,\tilde s.
\end{equation}
The latter transition in this sequence is due to the t-channel $W$
exchange
$$
u  \,\tilde d \to W^+ \to u \, \tilde s.
$$
The contribution to the $\eta - K^0$ matrix element from the
process~ (4) is determined by
\begin{multline}
\label{5} M_5^d = \langle K^0|O_5^d|\eta \rangle = \langle
K^0|(\bar d_R\gamma_\mu \lambda^a d_R)(\bar
d_L\gamma^\mu \lambda^a s_L)|\eta \rangle \\
= - (32/9) \langle K^0|(\bar d_R s_L)(\bar d_L  d_R)|\eta \rangle
= - (32/9) \langle K^0|\bar d_R s_L |0 \rangle \langle 0 | \bar
d_L d_R)|\eta \rangle \\
= - (32/9)(1/4) \langle K^0|\bar d \gamma_5 s |0 \rangle \langle
0 | \bar d \gamma_5   d)|\eta \rangle.
\end{multline}
Here $ \gamma_\mu , \gamma_5$ are the Dirac matrices and
$\lambda^a$ are the Gell - Mann colour matrices. In deducing these
relations the Fierz identities have been used for both the Dirac
and Gell - Mann matrices. Besides, the terms forbidden by the
colour-conserving law have been omitted.

Each matrix element in formula~ (5) can further be expressed in
terms of the $K_{l2}$ and $\pi_{l2}$ decay constants in the
following way:
\begin{equation}
\label{6}
 \langle 0 | \bar d \gamma_5   d)|\eta \rangle
= \frac{p^{\,\mu}}{2m_d}  \langle 0 | \bar d \gamma_\mu \gamma_5
d)|\eta \rangle = \frac {1}{\sqrt 6} \, f_\eta \frac
{m^2_\eta}{2m_d} \, \phi_\eta,
\end{equation}
and analogously,
\begin{equation}
\label{7}
 \langle K^0|\bar d \gamma_5  s |0 \rangle = f_K \frac {m^2_\eta}{m_d + m_s}\, \phi_K^*,
\end{equation}
where $f_K,  f_\eta, $ and $\phi_K$ , $\phi_{\eta}$ are the decay
constants and wave functions of the mesons, respectively; $p$ is
total momentum of the system; $m_ d,  m_s $ and $m_\eta$ are
masses quarks and $\eta$ meson.   In these relations, the quark
models of the $\eta$ and $K^0$ mesons have been taken into
account: $\eta = \frac{1}{\sqrt 6} (u \tilde u + d \tilde d - 2s
\tilde s )$ and $K^0 = d \tilde s$, and consequently $p = p_q +
p_{\tilde q}$.

Now a set of equations (3), (5), (6), and (7) leads to the
following formula for the off - diagonal mass matrix element of
the Lagrangian:
\begin{equation}
\label{8} \delta \left(m^2 \right) = - \frac {2c_5 G \sin{2\theta}
f_K  f_\eta m_{\eta}^4}{9\sqrt 3\, m_d (m_d + m_s)} \, \phi_K^*
\phi_\eta,
\end{equation}

In this case, the states $ |\eta \rangle $ and $ |K^0 \rangle $
 (with strangeness $ S = 0 $ and 1) are not stationary ones and are
  actually coherent superposition of different energy eigenstates:
\begin{eqnarray}
|\eta \rangle &=& |1 \rangle \cos{\varphi} +
|2 \rangle \sin{\varphi},  \nonumber \\
|K^0 \rangle &=& -|1 \rangle \sin{\varphi} + |2 \rangle
\cos{\varphi},  \label{9}
\end{eqnarray}
where the vectors $ |1\rangle $ and $ |2\rangle $ correspond
 to eigenstates with energies $ E_1 $ and $ E_2 $, $ \varphi $ is
 a mixing angle. In basis $ |1\rangle $, $ |2\rangle $ the mass matrix
 is diagonal. As to an initial basis $ |\eta\rangle $, $ |K^0\rangle $,
 the mass matrix contains the off-diagonal matrix terms $ \delta (m^2) $
in this representation. The mixing angle $ \varphi $ and $\delta
(m^2) $
 are related by
\begin{equation}
\label{10}
 \tan{2\varphi} = \frac {2 \,\delta (m^2)}{m_K^2 - m_{\eta}^2}.
\end{equation}

The new eigenstates $ |1\rangle $ and $ |2\rangle $ have different
 evolution properties, therefore, a strange content of a meson beam
 proves to be time dependent. Consequently, an initially pure
$ \eta $ beam will spontaneously generate $ K^0 $ mesons.
Combining equations (9) with the stationary state evolution law
gives the following formula for the probability of finding $ K^0
$ at time $ t $ :
\begin{multline}
\label{11}
 P(t)= \left|\langle K^0|\psi(t)\rangle \right|^2 \,
  = \, \sin^2{\varphi} \cos^2{\varphi} \left|e^{-i(E_1 - i\frac{\Gamma_1}{2})t}
- e^{-i(E_2 - i\frac{\Gamma_2}{2})t}\right|^2 \\
= \frac{1}{4} \sin^2{2\varphi} \left(e^{-\Gamma_1 t} +
e^{-\Gamma_2 t} - 2e^{-\frac{\Gamma_1 + \Gamma_2}{2} t} \cos(E_2
- E_1)t \right),
\end{multline}
where $ E_1 \approx E_{\eta},\quad \Gamma_1 = \Gamma_{\eta} \, $
and \quad $ E_2 \approx E_K, \quad \Gamma_2 = \Gamma_K $. \quad
$\Gamma$ is decay width: $\Gamma_{\eta} = 1/\tau_{\eta}$ and
$\Gamma_K = 1/\tau_K $.

The quantities, involved in the formula (11), have the following
values at rest [13]: \quad $|E_2 - E_1| = |m_K - m_{\eta}|
\approx ~ 50$ \, MeV,
 $\quad \Gamma_1 = \Gamma_{\eta} = 1.30$ \, keV ,\quad and $\Gamma_2
 = \Gamma_K  \approx 0 $; \quad hence, \quad $|E_2 - E_1| \gg \Gamma_1
\gg \Gamma_2 $. It means that the function (11) oscillates with
large frequency, and only its average value makes sense:
\begin{equation}
\label{12} \langle P \rangle = \frac 12 \sin^2{2\varphi}.
\end{equation}
This expression is correct for the time intervals $t \ll
\tau_{\eta}, \tau_K $.

By means of (2), (8), (10), and (12)
 we obtain the accomplished formula for the $ p\,K^0 $ branching ratio:
\begin{equation}
\label{13} \frac {\Gamma(N^{*+} \to p\,K^0 )}{\Gamma_t}
=\frac{8c_5^2 G^2 \sin^2{2\theta} f_K^2 f_\eta ^2 m_\eta ^8} {243
m_d ^2 (m_d + m _s)^2(m_\eta ^2 - m_K ^2)^2} \frac {\Gamma(N^{*+}
\to  p\,\eta ) }{\Gamma_t}.
\end{equation}

\section{Numerical estimates}
\label{nume}

 Now we may proceed to numerical estimates. We assume that the constant
 $f_\eta$ is to be approximately equal to the constant of $f_\pi$.
 In turn, values of the $f_\pi$ and $f_K$  constants are extracted from the
experimental data about the $\pi_{l2}$ and $K_{l2}$ decays ~[15]:
$f_\eta  \approx f_\pi = 130 $ MeV and $f_K / f_\pi = 1.27$. In
accordance with the theoretical estimates ~[14,~ 15], the
parameter $c_5 = - 0.14 $.  Other quantities in formula (13) are
well-known ~[13]: $G = 1.17 \times 10^{-5}~ GeV^{-2},~ \theta =
13^{\circ},\, m_{\eta} = 548~ MeV,\quad  m_K = 498~ MeV,\quad
$~and~ $\Gamma(N^*N\eta)/\Gamma_t = 0.53$. The substitution of
these values in (13) gives
$$
\Gamma(N^{*+} \to p\,K^0)/\Gamma_t \approx 5 \times 10^{-11}.
$$

This result is due to the contribution from the main matrix
element $M_5^d$. As far as the neglected effects are concerned we
confine their consideration to a few comments.

The contribution to $\delta \left(m^2 \right)$ from the matrix
element  $M_5^s$  is reduced with respect to  $M_5^d$ by a factor
of $m_d/m_s \approx 1/20 $; the matrix element $M_5^u = 0$;
 and the matrix element $M_6^d = \langle K^0|(\bar d_R \gamma_\mu d_R)
 (\bar d_L\gamma^\mu  s_L)|\eta \rangle $ gives rise relative contribution
 which is not in excess of 6 \%. Other effects turn out to be much less important
 than those discussed above.

 As for width of the peak, the $\eta - K^0$ oscillation also forms
 a complementary resonance peak, peculiar to the exotic decays, which
 overlaps the $N^*(1535)$ resonance. As is obvious from the formula (11),
 its width should be approximately equal to the $\eta$ meson decay width
$\Gamma_{\eta} = 1.30$~ keV.

\section{Concluding remarks}
\label{con}

 We have interpreted here an emergence of the anomalous states $pK^0$
 and $n\,K^+$ as a result of the $\eta \rightleftarrows K^0$ oscillation
 which proved  to be an extremely effective process. This is due to two causes.
 First, the contribution to the non-diagonal  mass matrix element
 $\delta \left(m^2 \right)$  from the "penguin" process (4) is enhanced
 by a factor as large as $\chi^2 = \left(m_\eta^ 2/2m_d(m_d + m_s)
\right)^2 \approx 1 \times 10^5$. Second, the $\eta - K^0$ mixing
is essentially enhanced too since the $K^0$ and $\eta$ levels are
close.

While the decay modes considered are solely of low-probability,
nevertheless, there is a possibility of their experimental
observation. It may be realized in the reactions involving, for
example, the low-energy transitions of $\gamma^0 \, N \to \rho^0
\, N \to N^*(1535) \to \eta \, N \to K \, N$ or $K \, \Lambda \to
N^*(1535) \to \eta \, N  \to  K  \, N$ etc. This is accounted for
by the following property of these reactions: they proceed when
the momentum (in c. m. s.) of scattering particles is close to
zero in the vicinity of the resonance. Then, according to the
Breit - Wigner formula, their cross sections should have
extremely large values near the resonance.

Thus, it seems to be very plausible that results of these
processes have already been observed at least in some of the
experiments searching for pentaquarks.

\end{document}